\documentstyle[12pt,epsfig]{article}
\begin{document}
\centerline{\Large\bf Corrections to the fine structure constant in
$D$-dimensional}
\vspace*{0.050truein}
\centerline{\Large\bf space from the generalized uncertainty principle}
\vspace*{0.050truein}
\centerline{Forough Nasseri\footnote{Email: nasseri@fastmail.fm}}
\centerline{\it Physics Department, Sabzevar University of Tarbiat
Moallem, P.O.Box 397, Sabzevar, Iran}
\centerline{\it Khayyam Planetarium, P.O.Box 769, Neishabour, Iran}
\begin{center}
(\today)
\end{center}

\begin{abstract}
In this letter we compute the corrections to the fine structure
constant in $D$-dimensional space. These corrections stem from the
generalized uncertainty principle. We also discuss in three-space
dimension.
\end{abstract}

\section{Introduction}
Recent observations of spectral lines of distant quasars have suggested
that the fine structure constant  $\alpha=e^2/(4 \pi \epsilon_0 \hbar c)$
may have been slightly smaller in the very early universe \cite{1}.
Although these claims are still tentative and rather controversial,
they have helped rekindle interest in Dirac's old idea \cite{2}
that the fundamental ``constants'' of physics may vary in time.

Varying fine structure constant has been studied in many papers,
for example see \cite{mota,sri,ben,pei,fla,bar,kaz,hc}.
The authors of \cite{loren} derived the formulae for the time
variation of the gravitational constant $G$ and of the fine structure
constant $\alpha$ in various models with extra dimensions and analyzed
their consistency with the available observational data for distant
supernovae, see also \cite{mbelek}.

On the other hand, it is of interest to define an effective fine structure
constant $\alpha_{eff}$ in terms of an effective Planck constant
$\hbar_{eff}$ so that we define
$\alpha_{eff} \equiv e^2/(4\pi \epsilon_0 \hbar_{eff} c)$.
In this letter we obtain $\hbar_{eff}$ based on the generalized
uncertainty principle.

Based on simple and general considerations with either Newtonian
gravitational theory or general relativity theory, the generalized
uncertainty principle has been obtained in \cite{adler}.
Explicit expressions for the generalized uncertainty principle in
extra dimensions are given and their holographic properties investigated
\cite{scard}.
According to \cite{book}, in $D$ spatial dimensions the dimensionless
constant of nature is proportional to
\begin{equation}
\label{0}
h^{2-D}e^{D-1}G^{(3-D)/2}c^{D-4}.
\end{equation}
As argued in \cite{book} our universe appears to possess a collection
of fundamental or natural units of mass, length and time which can be
constructed from the physical constants $G$, $h$ and $c$. A dimensionless
constant can only be constructed if the electron charge, $e$, is also
admitted and then we obtain the dimensionless quantity $e^2/(hc)$,
first emphasized in \cite{somm}. In a world with $D$ spatial dimensions
the units of $h$ and $c$ remain $M L^2 T^{-1}$ and $L T^{-1}$ in mass (M),
length (L) and time (T). The units of $G$ become $M^{-1} L^D T^{-2}$.
Gauss' theorem relates $e$ to the spatial dimension and the units of
$e^2$ are $M L^D T^{-2}$ because the electric force changes in accord with
$F=q q'/r^2$. It is important to note that the system of units that has
been used in the above argument is not the international system
of units which we use in this letter.
As we know, in the international system of units, the fine structure
constant is defined $\alpha={e^2}/(4\pi\epsilon_0 \hbar c)$
and in the Heaviside-Lorentz system $\alpha={e^2}/(\hbar c)$.

We show here that when one considers the generalized uncertainty
principle, the corrections to the fine structure constant in
$D$-dimensional space can be drawn.

The plan of this paper is as follows.
In section 2, we review to obtain the fine structure constant in
$D$-dimensional space. In section 3, we present the
generalized uncertainty principle in $D$-dimensional space
and then obtain the corrected fine structure constant in $D$-dimensional
space due to the generalized uncertainty principle.
Finally, we discuss and conclude in section 4.

\section{Fine structure constant in $D$-dimensional space}

In this section, we derive the fine structure constant in $D$-dimensional
space, see \cite{book}.
First of all, we need to use the Newton's gravitational constant
in $D$-dimensional space. Let us first derive the exact relationship
between Newton's gravitational constants $G_D$ and $G_3=G$ in $D$
and $3$-dimensional space. Using the force laws in $D$ and
$3$-dimensional space, which are
defined by
\begin{eqnarray}
\label{13}
F_D=G_D\frac{m_1 m_2}{r^{D-1}},\\
\label{14}
F_3=G_3\frac{m_1 m_2}{r^2},
\end{eqnarray}
and the $D$-dimensional space Gauss' law, one can derive the exact
relationship between the gravitational constants $G_D$ and $G_3$
\begin{equation}
\label{15}
G_3=\frac{\Omega_{D-1}}{4\pi}\frac{G_D}{V_{(D-3)}},
\end{equation}
where $V_{(D-3)}$ is the volume of $(D-3)$ extra spatial dimensions
and $\Omega_{D-1}$ is the area of a unit $(D-1)$-sphere
\begin{equation}
\label{16}
\Omega_{D-1}=\frac{2 \pi^{D/2}}{\Gamma \left( \frac{D}{2} \right)}.
\end{equation}
For $D=3$, we have $\Omega_2=4\pi$ which is the surface area of
a unit $2$-sphere.
To obtain a general formula for the fine structure constant in
$D$-dimensional space we propose that in $D$-dimensional space the
fine structure constant can be written as
\begin{equation}
\label{17}
\alpha_D= \hbar^{\,\beta} e^{\,\gamma}
[\Omega_{D-1} {\epsilon_{D,0}}]^{\,\eta}
c^{\,\xi} G_D^{\,\tau},
\end{equation}
where ${\epsilon_{D,0}}$ is the permittivity constant of the vacuum in
$D$-dimensional space and its units can be obtained by
the electric force law in $D$-dimensional space
\begin{equation}
\label{18}
F_D=\frac{1}{\Omega_{D-1} {\epsilon_{D,0}}} \frac{q_1 q_2}{r^{D-1}}.
\end{equation}
From Eq. (\ref{18}) we obtain the units of ${\epsilon_{D,0}}$ to be
equal to $Q^2 M^{-1} L^{-D} T^2$.
In a world with $(D+1)$-spacetime the units of $\hbar$, $c$ and $e$ remain
$ML^2T^{-1}$, $LT^{-1}$ and $Q$ in Mass $(M)$, length $(L)$, time $(T)$
and electric charge $(Q)$. Using Eq. (\ref{15}) one can obtain the units of
the Newton's gravitational constant in $D$-dimensional space $G_D$ to be
equal to $M^{-1} L^D T^{-2}$.
Thus in $D$-dimensional space the dimensionless constant of nature,
i.e. the fine structure constant has the units of
\begin{equation}
\label{19}
\left( M L^2 T^{-1} \right)^{\beta} Q^{\gamma}
\left( Q^2 M^{-1} L^{-D} T^{2} \right)^{\eta}
\left(L T^{-1} \right)^{\xi}
\left( M^{-1} L^{D} T^{-2} \right)^{\tau}.
\end{equation}
Because the fine structure constant in $D$-dimensional space
is a dimensionless quantity, the sum of powers of mass ($M$),
length ($L$), time ($T$), electric charge ($Q$) must be vanished.
Therefore we have
\begin{eqnarray}
\label{20}
\beta-\eta-\tau&=&0,\\
\label{21}
2\beta-D\eta+\xi+D\tau&=&0,\\
\label{22}
-\beta+2\eta-\xi-2\tau&=&0,\\
\label{23}
\gamma+2\eta&=&0.
\end{eqnarray}
We know that in $3$-dimensional space we have
$\alpha=e^2/(4\pi \epsilon_0 \hbar c)$. So for $D=3$ we have
the following conditions
\begin{eqnarray}
\label{24}
\beta&=&-1,\\
\label{25}
\gamma&=&2,\\
\label{26}
\eta&=&-1,\\
\label{27}
\xi&=&-1,\\
\label{28}
\tau&=&0.
\end{eqnarray}
Using Eqs. (\ref{20})-(\ref{23}) and conditions (\ref{24})-(\ref{28})
one can obtain
\begin{eqnarray}
\label{29}
\beta&=&2-D,\\
\label{30}
\gamma&=&D-1,\\
\label{31}
\eta&=&\frac{1-D}{2},\\
\label{32}
\xi&=&D-4,\\
\label{33}
\tau&=&\frac{3-D}{2}.
\end{eqnarray}
Thus in $D$-dimensional space the fine structure constant is equal to
\begin{equation}
\label{34}
\alpha_D= \hbar^{2-D} e^{D-1} [\Omega_{D-1} {\epsilon_{D,0}}]^{(1-D)/2}
c^{D-4} G_D^{(3-D)/2}.
\end{equation}
This equation is a general formula for the fine structure constant
in $D$-dimensional space.
For $D=3$, Eq. (\ref{34}) leads us to
$\alpha=e^2/(4\pi \epsilon_0 \hbar c)$.

\section{Corrections to the Fine Structure Constant}
In $D$-dimensional space, the Heisenberg uncertainty principle is
written as
\begin{equation}
\label{1}
\Delta x_i \Delta p_j \geq \hbar \delta_{ij},
\end{equation}
where $x_i$ and $p_j$, $i,j=1...D$, are the spatial coordinates and
momenta, respectively.
In $D$-dimensional space, the maximum uncertainty in the position of
an electron in the ground state in hydrogen atom
is equal to the radius of the first Bohr orbit, $r_B$,
\begin{equation}
\label{3}
\Delta x=r_B= \left(
\frac{\Omega_{D-1} \epsilon_{D,0} \hbar^2}{m e^2} \right)^{\frac{1}{4-D}},
\end{equation}
where $m$ is the mass of the electron.
Eq.(\ref{3}) in the case of $D=3$ yields the
radius of the first Bohr orbit, so-called Bohr radius,
$r_B=4 \pi \epsilon_0 \hbar^2/(m e^2)=5.29 \times 10^{-11} m$.
The general form of the generalized uncertainty
principle is
\begin{equation}
\label{5}
\Delta x_i \geq \frac{\hbar}{\Delta p_i}+{\hat \beta}^2 L_P^2
\frac{\Delta p_i}{\hbar},
\end{equation}
where $L_P=(\hbar G_D/c^3)^{1/(D-1)}$ is the Planck length and
$\hat \beta$ is a dimensionless constant of order one.
There are many derivations of the generalized uncertainty principle,
some heuristic and some more rigorous. Eq. (\ref{5}) can be derived
in the context of string theory and non-commutative quantum mechanics.
The exact value of ${\hat \beta}$ depends on the specific model.
The second term in r.h.s of Eq.(\ref{5}) becomes effective when momentum
and length scales are of the order of Planck mass and of the Planck
length, respectively. This limit is usually called quantum regime.
From Eq.(\ref{5}) we solve for the momentum uncertainty in terms of
the distance uncertainty, which we again take to be the radius of 
the first Bohr orbit.
From Eq.(\ref{5}), we are led to the following momentum uncertainty
\begin{equation}
\label{6}
\frac{\Delta p_i}{\hbar}=\frac{\Delta x_i}{2 {\hat \beta}^2 L_P^2}
\left ( 1- \sqrt{1- \frac{4 {\hat \beta}^2 L_P^2}{{\Delta x_i}^2}} \right).
\end{equation}
Eq.(\ref{3}) gives
\begin{equation}
\label{7}
\frac{\Delta x}{L_P}=\left( \frac{\Omega_{D-1} \epsilon_{D,0}
M_P^3 G_D}{m e^2} \right)^{\frac{1}{4-D}},
\end{equation}
where $M_P=[\hbar^{D-2}/(c^{D-4}G_D)]^{1/(D-1)}$ is the Planck mass.
Recalling the standard uncertainty principle
$\Delta x_i \Delta p_i \geq \hbar$, we define an ``effective'' Planck
constant $\Delta x_i \Delta p_i \geq h_{eff}$.
From Eq.(\ref{5}), we can write
\begin{equation}
\label{8}
\Delta x_i \Delta p_i \geq \hbar \left[ 1 + {\hat \beta}^2 L_P^2
\left( \frac{\Delta p_i}{\hbar} \right)^2 \right].
\end{equation}
So we can define the effective Planck constant as
\begin{equation}
\label{9}
\hbar_{eff}=\hbar \left[ 1 + {\hat \beta}^2 L_P^2
\left( \frac{\Delta p_i}{\hbar} \right)^2 \right].
\end{equation}
From Eqs.(\ref{6},\ref{7}) and (\ref{9}) we get
\begin{eqnarray}
\label{10}
\hbar_{eff}&=&\hbar \Bigg[ 1+\frac{1}{4 {\hat \beta}^2}
\left("\frac{\Omega_{D-1} \epsilon_{D,0} M_P^3 G_D}{m e^2}
\right)^{\frac{2}{4-D}}\nonumber\\
&\times& \left( 1 -
\sqrt{ 1 - 4 {\hat \beta}^2
\left( \frac{m e^2}{\Omega_{D-1} \epsilon_{D,0} M_P^3 G_D}
\right)^{\frac{2}{4-D}} }
\right)^2 \Bigg].
\end{eqnarray}
Because the value of the expression
$
\frac{m e^2}{\Omega_{D-1} \epsilon_{D,0} M_P^3 G_D}
$
is much less than one, we can expand Eq.(\ref{10}). Therefore, we have
\begin{equation}
\label{11}
\hbar_{eff} \simeq \hbar \left[ 1 + {\hat \beta}^2
\left( \frac{m e^2}{\Omega_{D-1} \epsilon_{D,0} M_P^3 G_D} \right )^{\frac{2}{4-D}} \right].
\end{equation}
So the effect of the generalized uncertainty principle can be taken
into account by substituting $\hbar_{eff}$ with $\hbar$.
For three-space dimension, from Eq.(\ref{11}) we obtain
\begin{equation}
\label{12}
\hbar_{eff} \simeq \hbar \left[ 1 + {\hat \beta}^2
\times 9.30 \times 10^{-50} \right].
\end{equation}
Because the factor $10^{-50}$ is much less than one,
we conclude that the value of the effective Planck constant
is very close to the value of the standard Planck constant.
Substituting the effective Planck constant $\hbar_{eff}$
from Eq.(\ref{10}) into Eq.(\ref{34}) we obtain the effective and
corrected fine structure constant due to the generalized uncertainty
principle
\begin{equation}
\label{35}
\alpha_{D,eff}= \hbar_{eff}^{2-D} e^{D-1}
[\Omega_{D-1} {\epsilon_{D,0}}]^{(1-D)/2} c^{D-4} G_D^{(3-D)/2}.
\end{equation}
This equation shows the corrections to the fine
structure constant from the generalized uncertainty principle
in $D$-dimensional space. For $D=3$, from (\ref{35}) we obtain
\begin{equation}
\label{36}
\alpha_{eff}= \hbar_{eff}^{-1} e^{2}
[4 \pi \epsilon_0]^{-1} c^{-1}=
\frac{e^2}{4 \pi \epsilon_0 \hbar_{eff} c}.
\end{equation}
From (\ref{12}), we are led to
\begin{equation}
\label{37}
\alpha_{eff} \simeq \frac{e^2}{4 \pi \epsilon_0 \hbar c}
\left[ 1 - {\hat \beta}^2
\times 9.30 \times 10^{-50} \right].
\end{equation}
This equation shows the corrections to the fine structure constant
in three-space dimension from the generalized uncertainty principle.

\section{Conclusions}
In this paper, we have examine the effects of the generalized uncertainty
principle in the fine structure constant in $D$-dimensional space.
We also discuss our calculations in three-space dimension.
The general form of the generalized uncertainty principle is given
by (\ref{5}). The Planck constant $\hbar$ undergoes corrections from the
generalized uncertainty principle, and changes into $\hbar_{eff}$ as given
in Eq.(\ref{10}) or approximately (\ref{11}). Then we obtain the
corrections to the fine structure constant in $D$-dimensional
space as Eq.(\ref{35}). As seen from Eq.(\ref{12}), the value of
$\hbar$ and $\hbar_{eff}$ is very close to each other because
the factor $10^{-50}$ is much smaller than one. For this reason
the value of the fine structure constant due to the generalized
uncertainty principle is very close to the value of the fine structure
constant in the standard uncertainty principle which is $\frac{1}{137}$.
The numerical results of Eq.(\ref{35}) are
in progress by the author.

\section*{Acknowledgments}
The author thanks Amir and Shahrokh for useful helps.

\end{document}